\documentstyle[12pt]{article}
\newcommand{\beq}{\begin{equation}}
\newcommand{\eeq}{\end{equation}}
\newcommand{\bea}{\begin{eqnarray}}
\newcommand{\eea}{\end{eqnarray}}
\newcommand{\D}{\displaystyle}
\newcommand{\U}{\hat{U}}
\input epsf

\begin{document}

\hfill \vbox{\hbox{UCLA/97/TEP/22}
             \hbox{COLO-HEP-392}
             \hbox{hep-lat/9711009}} 
\begin{center}{\Large\bf Vortices and Confinement at 
weak coupling}\\[2cm] 
{\bf Tam\'as G. Kov\'acs}\footnote{Research supported by 
DOE grant DE-FG02-92ER-40672} \\
{\em Department of Physics, University of Colorado, Boulder 
CO 80309-390}\\
{\sf e-mail: kovacs@eotvos.Colorado.EDU}\\[5mm] 

and\\[5mm]

{\bf E. T. Tomboulis}\footnote{Research supported by 
NSF grant NSF-PHY 9531023}\\
{\em Department of Physics, UCLA, Los Angeles, 
CA 90095-1547}\\
{\sf e-mail: tombouli@physics.ucla.edu}
\end{center}
\vspace{1cm}

\begin{center}{\Large\bf Abstract}
\end{center}
We discuss the physical picture of thick vortices  
as the mechanism responsible for confinement at arbitrarily 
weak coupling in $SU(2)$ gauge theory. By introducing 
appropriate variables on the lattice we distinguish between 
thin, thick and `hybrid' vortices, the latter involving 
$Z(2)$ monopole loop boundaries. We present numerical 
lattice simulation results that demonstrate that the 
full $SU(2)$ string tension at 
weak coupling arises from the presence of vortices linked 
to the Wilson loop. Conversely, excluding linked vortices 
eliminates the confining potential. The numerical results 
are stable under alternate choice of lattice action as well 
as a smoothing procedure which removes short distance 
fluctuations while preserving long distance physics.

\vfill
\pagebreak     
\section{Introduction} 

Arguments for the presence of spread-out tubes 
of color-magnetic flux, thick `vortices', being the 
essential feature responsible for maintaining confinement 
at arbitrarily weak coupling in $SU(N)$ gauge theory 
were expounded some time ago \cite{MP} - \cite{T1}. A key idea is 
that such extended structures cost very little action 
locally, and thus are not directly suppressed at large 
$\beta$. By gradual variation of the gauge fields, they can 
disorder the vacuum over long scales.
The infrared physical picture is, of course, independent of 
any ultraviolet cutoff details, but, as always 
with such nonperturbative questions, mathematically precise 
formulations have been possible only on the lattice. 
Thick vortices form closed extended structures which 
are topologically characterized, 
in the continuum extrapolation, by $\pi_1(SU(N)/Z(N))=Z(N)$. 
`Punctured' thick vortices, whose (small) `hole' boundary 
is a Dirac monopole current loop, are also possible and 
survive at large $\beta$ \cite{T1}.
(These Dirac monopoles are also classified by the 
non-trivial elements of $\pi_1(SU(N)/Z(N))$.)
In the $SU(N)$ lattice gauge theory there also occur 
`thin' vortex excitations of the $Z(N)$ part of 
the group. These are localized to one lattice spacing 
thickness, and hence are sensitive to the short distance 
details such as the precise choice of the plaquette action.
Long thin vortices are very efficient at 
disordering the system at strong coupling, but are 
energetically heavily suppressed and become irrelevant 
at large $\beta$.    

In this paper we first discuss in detail the various vortex 
excitations possible in the $SU(N)$ lattice gauge theory 
(section 2). We treat explicitly the simplest $N=2$ case 
since no additional physical features appear in the general 
$N$ extension which is straightforward. The proper distinction 
between thin and thick vortices and their interactions 
seems to have occasioned some confusion in the literature. 
A clean separation can be achieved by an exact rewriting  
of the $SU(2)$ theory in terms 
of $SU(2)/Z(2)\sim SO(3)$ and $Z(2)$ variables  
\cite{T}, \cite{MP}. We eschew any mathematical derivations that 
can be extracted from the literature, and give a detailed 
physical discussion of the various excitations and their 
interactions.\footnote{This is an extended version of the argument 
given in \cite{T1}.} 
In section 3 we examine the Wilson loop and its interaction 
with vortices. The expression for the Wilson loop operator 
in the $SO(3)$-$Z(2)$ formalism shows that it is essentially a 
vortex counter. The fluctuation of the operator between 
positive and negative values is determined {\it solely} by the 
number (mod 2) of vortices linked with the loop. This then 
allows us to examine the vortex contribution to the string 
tension numerically (section 4). The 
string tension extracted from the full Wilson loops was 
compared to the same quantity extracted 
from the expectation of only the sign fluctuation counting the 
linking of the vortices. The computation was first performed  
with the Wilson action, and then also with a fixed point 
action in conjunction with a `smoothing' procedure
based on the renormalization group. 
The point of performing the comparison 
also under smoothing is that smoothing removes short distance 
fluctuations while preserving long distance physics. 
In particular, the string tension of the full Wilson loop 
remains unchanged under the smoothing procedure. A necessary 
test then of any claim concerning the long distance physics 
is that it remain invariant under the smoothing procedure.    
The numerical results demonstrate that 
the confining potential arises from the presence of the 
vortices linked to the loop: the full string tension is, 
remarkably, reproduced from the expectation of the vortex 
counting sign. Conversely, allowing no (mod 2) 
vortices to link with the loop eliminates the confining 
potential. Closely related results have been reported 
in \cite{G}. Our conclusions are presented in section 5.

\section{Vortices - thin, thick and hybrid}

For $SU(N)/Z(N)$ gauge fields in the continuum, vortices 
are topologically classified by $\pi_1(SU(N)/Z(N))=Z(N)$. This 
means that a vortex nontrivially linked to a Wilson loop (trace 
of the parallel transport matrix of the gauge field connection),  
taken in the fundamental representation of the covering group 
$SU(N)$, contributes a factor $z \in Z(N)$, $z\neq 1$. A vortex 
forms a closed 2-dim surface in $d=4$ (a loop in $d=3$) so that 
it links with a Wilson loop $C$ if it pierces once any 
surface bounded by $C$. Topologically, it is also possible 
to have gauge field configurations representing `open' 
vortices (Dirac sheets). The boundary of an open 2-dim vortex 
sheet represents a monopole loop (monopole-antimonopole 
pair in 3-dim). These are Dirac monopoles, also classified 
by the non-trivial elements of $\pi_1(SU(N)/Z(N))$ \cite{CO}.  

Now in the continuum, where the gauge field is an element of 
the Lie group algebra, there is no local distinction between 
the pure $SU(N)/Z(N)$ and $SU(N)$ gauge theories. In the lattice 
formulation, in terms of group element bond variables, of 
course there is. The two differ by the dynamics  
of the additional $Z(N)$ degrees of freedom present in the 
$SU(N)$ case. Exciting these Z(N) degrees of freedom 
on a stack of plaquettes forming a 2-dim closed wall ( a closed 
loop in $d=3$) gives a `thin' $Z(N)$ vortex. These are of course 
the vortices already present in a pure $Z(N)$ lattice gauge 
theory (LGT). They are `thin' because they necessarily have 
thickness of one lattice spacing. At small $\beta$, they 
are very efficient at disordering the vacuum. At large $\beta$, 
however, they are heavily suppressed by the $SU(N)$ 
plaquette action, and get progressively frozen out as $\beta$ 
increases. Correspondingly, the pure $Z(N)$ LGT gets into a 
Higgs phase; whereas the distinction between the $SU(N)$ and 
$SU(N)/Z(N)$ LGT disappears, as it should, as the continuum 
limit is approached. Thus it is only the {\it non-Abelian} 
dynamics of the lattice analogs of the topological $Z(N) = 
\pi_1(SU(N)/Z(N))$ vortices, which can, 
if at all, affect the large $\beta$ long distance dynamics. 

The lattice literature contains many confused or incorrect 
statements due to failure to properly distinguish between 
the excitations of the $Z(N)$ part versus 
the (lattice analogs of the) topological 
$\pi_1(SU(N)/Z(N))=Z(N)$ excitations of the 
$SU(N)/Z(N)$ part of the $SU(N)$ gauge group, and their 
respective energetics. A formalism that allows one to 
accomplish such a separation cleanly introduces separate 
$SU(N)/Z(N)$ and $Z(N)$ variables \cite{T},\cite{MP}. 
It is important, of course, that this is done in a 
gauge-invariant manner, and gives an exact rewriting of 
the partition function and all observables in terms of 
the new variables.        

From now on we restrict to $SU(2)$, which is the actual 
case of our numerical simulations below. The extension 
to any $N$ is straightforward. Consider then the standard 
$SU(2)$ theory partition function on a lattice $\Lambda$
\beq
Z_\Lambda = \int\;\prod_b\,dU_b\;\exp\left( \sum_p\,\beta 
{\rm tr}U_p\right) \quad,\label{PF}
\eeq
where, as usual, we wrote $U_p=\prod_{b\in p}
U_b$ for the product of bond variables $U_b$ around the 
plaquette $p$. 

We now introduce new $Z(2)$ variables $\sigma_p \in 
\{\pm1\}$ residing on plaquettes. We write 
$\sigma_c \equiv \prod_{p\in  c}\sigma_p$ for 
the product of the $\sigma_p$'s around the faces of 
the cube $c$. 
We also introduce the coset bond variables $\hat{U}_b 
\in SU(2)/Z(2)\sim SO(3)$. The configuration space of the  
the $SU(2)$ bond variables on the lattice $\Lambda$ is split 
into equivalence classes, each class corresponding to one 
coset bond variable configuration $\{\hat{U}_b\}$ on $\Lambda$. 
Thus two $SU(2)$ configurations $\{U_b\}$ and $\{U_b^\prime\}$ 
on $\Lambda$ are representatives of the same coset 
configuration $\{\hat{U}_b\}$ if and only if one has 
$U^\prime_b = U_b\gamma_b$, for some $\gamma_b \in Z(2)$, for 
every bond $b$ on $\Lambda$. Now given a coset configuration 
$\{\U_b\}$, pick a representative $\{U_b\}$ and 
let $\eta_p \equiv \mbox{sign}\:U_p$. Then the quantity 
\beq
\eta_c(\hat{U}) \equiv \prod_{p\in  c}\;\eta_p 
\quad ,\label{eta}
\eeq    
the product of $\eta_p$ around the faces of a cube $c$, 
depends, as indicated, only on the coset variables since 
it is invariant under $U_b \to U_b\gamma_b$ for 
$\gamma_b\in Z(2)$. In other words, it is independent of 
the representative used to compute it.  

Now one can show \cite{MP}, \cite{T} that (\ref{PF}) can be 
written in the form 
\beq
Z_\Lambda = \int\;\prod_b\,dU_b\,\prod_p\,d\sigma_p\;
\prod_c\,\delta[\eta_c\sigma_c]\;\exp\left(\,\beta|{\rm tr}
U_p|\sigma_p\right)\quad.\label{PF1}
\eeq
In (\ref{PF1}), and what follows, the `delta function' on 
$Z(2)$ simply stands for  
\beq
\delta(\tau) \equiv \frac{1}{2}[1 + \tau], \qquad\quad 
\tau\in Z(2)\quad,
\eeq 
so that $\delta(\tau) =1$ for $\tau=1$, $\delta(\tau)=0$ for $\tau 
=-1$. Also, $\int\;d\sigma_p (\cdots) \equiv \sum_{\sigma_p
=\pm 1}(\cdots)$ stands for `integration' over the discrete 
$Z(2)$ group.  

The crucial point is that the integrand in (\ref{PF1}) 
depends only on the $\hat{U}$'s since it is invariant under 
the local transformation $U_b\to U_b\gamma_b$ 
for arbitrary $\gamma_b\in Z(2)$. 
In particular, the action becomes the product of a 
$Z(2)$ part and an $SO(3)$ part. The $Z(2)$ part, 
which is given simply 
by the plaquette variable $\sigma_p$, determines the 
sign of the action. The $SO(3)$ part is non-negative. 
In this connection note the relation $|{\rm tr}U|^2 = 
|\chi_{1/2}(U)|^2 = 1+\chi_1(U) = 1+\chi_1(\hat{U})$, 
where $\chi_{1/2}(U)$ and $\chi_1(U)$ denote the fundamental 
and adjoint representation characters of 
$SU(2)$. Thus the $U$-integration 
is in fact a $\U$-integration, i.e. in 
(\ref{PF1}) one has
\beq  
\prod_b\;dU_b = {\rm Constant}\times\prod_b\;d\U_b
\quad,
\eeq 
where $d\U_b$ is the Haar measure over $SO(3)$. 

Similarly, for the expectation of a Wilson loop one finds: 
\bea
W[C]&=& \frac{1}{\D Z_\Lambda} \int\;\prod_b\,dU_b\;{\rm tr}U[C]
\,\exp\left( \sum_p\,\beta {\rm tr}U_p\right)\label{W}\\
  &=& \frac{1}{\D Z_\Lambda} 
\int\;\prod_b\,dU_b\,\prod_p\,d\sigma_p\;\prod_c\;
\delta[\eta_c\sigma_c]\;{\rm tr}U[C]\:\eta_S\:\sigma_S \;
\exp\left(\,\beta|{\rm tr}U_p|\sigma_p\right)  
\quad.\label{W1}
\eea 
In (\ref{W1}), $U[C]=\prod_{b\in C}\:U_b$ stands for the 
product of the $U_b$'s around the loop $C$, and we 
introduced the notations:
\beq
\eta_S\equiv \prod_{p\in S}\;\eta_p \qquad,\qquad 
\sigma_S\equiv \prod_{p\in S}\;\sigma_p \quad,
\label{S}
\eeq
where $S$ is a surface bounded by the loop $C$, i.e. $C=
\partial S$. 
It is easily seen that (\ref{W1}) does 
not depend on the choice of surface $S$. Furthermore, the 
quantity ${\rm tr}U[C]\,\eta_S$ depends only on the $SO(3)$ 
bond variables $\U_b$, since it is invariant under 
$U_b \to U_b\gamma_b$, $\gamma_b \in Z(2)$. The Wilson 
loop operator is thus expressed as a product of an $SO(3)$ 
and a $Z(2)$ factor. Similarly, any other observable, 
such as a t'Hooft loop, or the electric and magnetic flux 
free-energies can be easily written in terms of the new 
variables.  

Equations (\ref{PF1}) and (\ref{W1}) then reexpress the $SU(2)$ LGT 
as a coupled $SO(3)$-$Z(2)$ theory. This rewriting is 
exact and gauge invariant. It is very convenient to evaluate 
all $SO(3)$ quantities in terms of $SU(2)$ representatives, 
as in (\ref{PF1}), (\ref{W1}).\footnote{They can always, of 
course, also be expressed directly in (a character 
expansion in) $SO(3)$ (integer spin) representations, as 
indicated above for the action plaquette function.}
At the risk of being 
repetitive, let us point out again that, once the passage to 
the $SO(3)$-$Z(2)$ formalism is made, the quantity 
$\eta_p = \mbox{sign}\:U_p$ \ for a representative of 
a $\{\U_b\}$ configuration has nothing to do with the 
sign of the action in the $SU(2)$ formalism (\ref{PF}),  
(\ref{W}). This sign, as already noted, is supplanted by 
the $Z(2)$ variable $\sigma_p$. The representative-dependent 
$\eta_p$'s can appear in physical quantities 
only in $SO(3)$ representative-independent combinations, 
as e.g. (\ref{eta}). 
The resulting expressions (\ref{PF1}), (\ref{W1}) of 
the $SO(3)$-$Z(2)$ formulation 
have a physically rather transparent 
form exhibiting the presence and manner of coupling of 
the various possible topological excitations in the LGT 
(\ref{PF}).  

\begin{figure}[ht]
\begin{minipage}{155mm}
{\ }\epsfbox{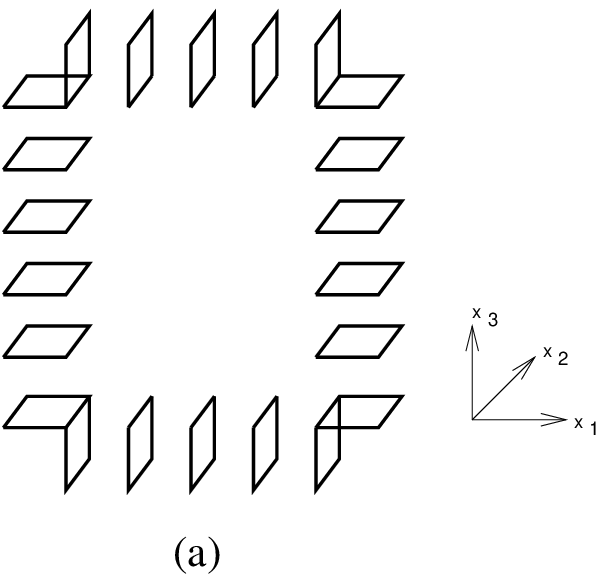}\hfill\epsfbox{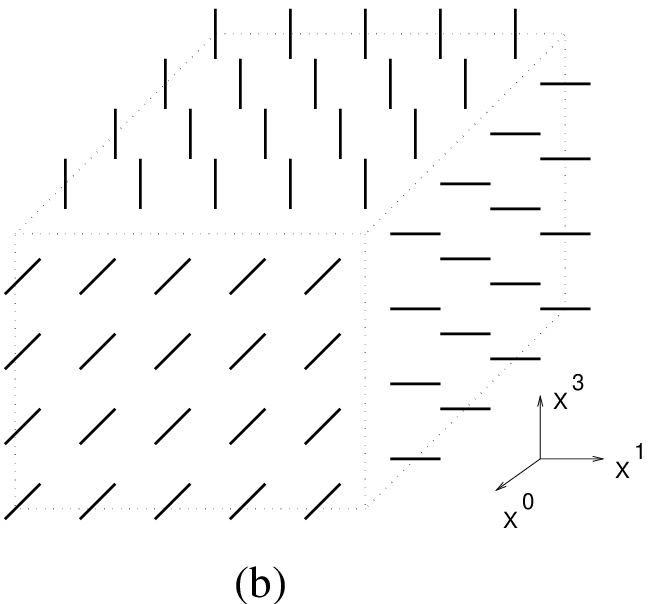}{\ }\\
\caption[Thin vortex]{\label{tv}Thin vortex: (a) d=3 or 
d=4 in [123]-section view; (b) d=4 in [130]-section
3-dimensional view. Plaquettes protruding in 
$x^2$-direction carry $\sigma_p=-1$.}
\end{minipage}
\end{figure}

Consider a configuration where $\sigma_p=1$ everywhere 
except on a stack of plaquettes forming a loop (Figure 
\ref{tv}(a)) where $\sigma_p=-1$. 
This is a $Z(2)$ vortex in 3 dimensions 
or a 3-dimensional section of a vortex in 4 dimensions. 
In $d=4$ there is an extra dimension to move in, so
by translation of the loop a vortex forms a 2-dimensional 
closed surface (Figure \ref{tv}(b)). The short lines in 
Figure \ref{tv}(b) represent a set of bonds. The 
plaquette protruding in the $x^2$-direction out 
of each of these bonds carries $\sigma_p=-1$. A 
3-dimensional $[\mu\nu\lambda]=[123]$ section 
gives then Figure \ref{tv}(a). These vortices, 
generated by the excitations of the $Z(2)$ 
$\sigma_p$ variables, are $Z(2)$ `thin' vortices, 
alluded to above . Note that, from (\ref{PF1}), the 
action cost for exciting such a $Z(2)$ vortex is 
directly proportional to the area of the vortex sheet. 

\begin{figure}[ht]
\begin{minipage}{155mm}
{\ }\epsfbox{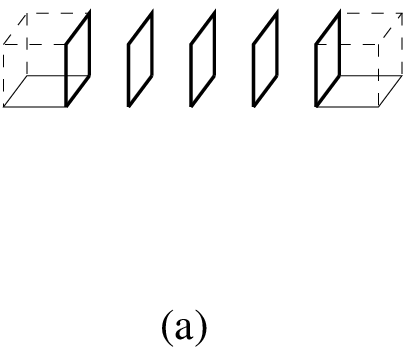}\hfill
\epsfbox{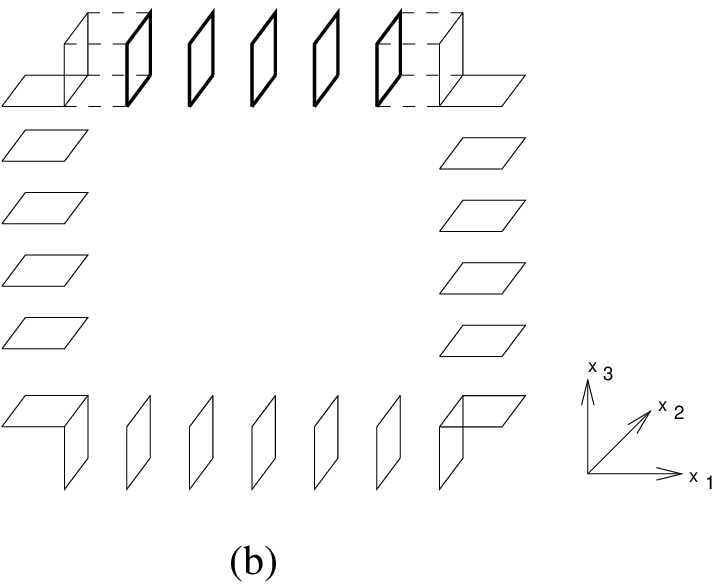}{\ }\\
\caption[Open vortex]{\label{ov} (d=3): (a)
$Z(2)$ monopole pair (cubes) joined by 
$\sigma$-string (open thin vortex); (b) the 
complete configuration including the $\eta$-string 
to form a `hybrid' vortex (see text).}
\end{minipage}
\end{figure}

Consider next opening the thin vortex by breaking the 
loop of $\sigma=-1$ plaquettes in Figure \ref{tv}(a) 
as depicted in Figure \ref{ov}(a). The two cubes at 
the two ends necessarily satisfy $\sigma_c=-1$. 
A cube with $\sigma_c=-1$ is the site of a $Z(2)$ 
monopole, and Figure \ref{ov}(a) depicts a 
monopole - antimonopole pair joined by an open thin 
vortex, i.e. a string of $\sigma_p=-1$ plaquettes 
carrying the $Z(2)$ flux. As just noted, there is a direct 
action cost associated with this $\sigma$-string. 

Now, because of the $Z(2)$ 
$\delta$-function constraint in the measure in (\ref{PF1}), 
a cube with $\sigma_c=-1$ must also have $\eta_c=-1$. 
A cube for  which $\eta_c=-1$ is the site of the lattice 
analog of a $\pi_1(SO(3))=Z(2)$ monopole. 
As pointed out above this statement depends only 
on the $SO(3)$ coset configuration $\{\U_b\}$ 
on the lattice, i.e. the presence or absence of such 
a monopole on a given cube is a gauge-invariant 
feature of each $SO(3)$ configuration. Any one 
representative $\{U_b\}$ 
of an $SO(3)$ $\{\U_b\}$ configuration with a monopole 
on a given cube will necessarily have 
a string of plaquettes, the Dirac string, 
beginning at the cube in question, on which 
$\eta_p=-1$. The string has to end at another monopole 
cube. Configurations with monopoles that contribute to 
the partition function then are of the form depicted in 
Figure \ref{ov}(b) in 3 dimensions; heavy plaquettes 
carry $\sigma_p=-1$, light plaquettes carry 
$\eta_p=-1$. Note that the location 
and shape of the string (light plaquettes)  
depends on the choice of representative; it can differ 
for different representatives of the same $\{\U_b\}$ 
configuration since 
it may be moved around at will, as a Dirac string should, by 
letting $U_b\to\gamma_bU_b$,\ $\gamma_b\in Z(2)$, i.e. 
change of representative. The 
$\eta$-string is then ``invisible'' to the $\{\U_b\}$ 
configuration on $\Lambda$, and hence to the measure, and 
in particular to the action in (\ref{PF1}), (\ref{W1}). 
There is no cost in action associated with the 
location of the Dirac $\eta$-string. 
\begin{figure}[htb]
\begin{minipage}{155mm}
{\ }\hfill\epsfysize=5cm\epsfbox{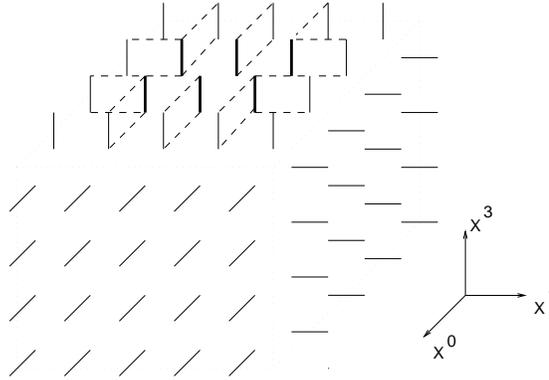}\hfill{\ }\\
\caption[Hybrid vortex]{\label{hv} Hybrid vortex (d=4) in
3-dim [130]-section view. Plaquettes 
protruding in the remaining $x^2$-direction out of 
light (heavy) bonds carry $\eta_p=-1$ ($\sigma_p=-1$);
the boundary between the light and heavy sets is 
a $Z(2)$ monopole current loop (see text).}
\end{minipage}
\end{figure}  

In $d=4$ the monopoles, i.e. the set of cubes on which 
$\eta_c=-1$, form closed monopole current 
loops reflecting magnetic current conservation. This follows 
directly from the definition of 
the quantity $\eta_c$, eq. (\ref{eta}).\footnote{Indeed 
it follows from (\ref{eta}) that $\eta_c$ obeys the 
identity
\[ \prod_{c\; \in\; h}\;\eta_c = 1 \quad ,\]
where the product is over all cubes $c$ forming the boundary 
of the elementary hypercube $h$. Geometrically, this 
means that the cubes on which $\eta_c=-1$ form closed sets 
on the dual lattice. In $d=4$, a cube is dual to a bond, 
so the cubes form closed loops of dual bonds.\label{F}} 
A string then sweeps out a Dirac sheet bounded 
by the corresponding monopole loop. So in $d=4$ the 
configuration in Figure \ref{ov}(b) gives rise to a 
`hybrid' vortex forming a closed 2-dimensional 
surface as shown in Figure \ref{hv}.
Here again, $\sigma_p=-1$ on the plaquette protruding in 
the $x^2$-direction out of every heavy bond, whereas 
$\eta_p = -1$ on the plaquette protruding in the 
$x^2$-direction out of every light bond shown. 
The plaquettes containing both a light and a 
heavy bond in their boundary are also shown in the figure. 
The cube protruding in the $x^2$-direction out of each of 
these plaquettes then has $\sigma_c= \eta_c = -1$. 
This set of cubes forms the monopole current 
loop (cp. footnote \ref{F}). A $[123]$-section 
view of Figure \ref{hv} gives then Figure \ref{ov}(b).  
Such a hybrid vortex may be viewed as put together by 
joining an `open' $\pi_1(SO(3))$ vortex plaquette sheet 
and an open $Z(2)$ vortex plaquette sheet along their 
respective monopole loop boundaries. These boundaries 
must coincide, as noted above, because open 
vortices as in Figure \ref{ov}(a) cannot exist 
due to the constraint in the measure in (\ref{PF1}).

\begin{figure}[hb]
\begin{minipage}{155mm}
{\ }\hfill\epsfysize=5cm\epsfbox{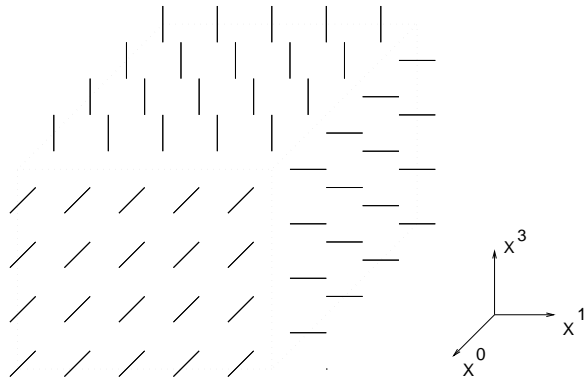}\hfill{\ }\\
\caption[Thick vortex]{\label{tckv} Thick vortex closed 
Dirac sheet (d=4) in 3-dim [013]-section view; plaquettes 
protruding in the remaining $x^2$-direction carry 
$\eta_p=-1$.}
\end{minipage}
\end{figure}

\begin{figure}[htb]
\begin{minipage}{155mm}
{\ }\epsfbox{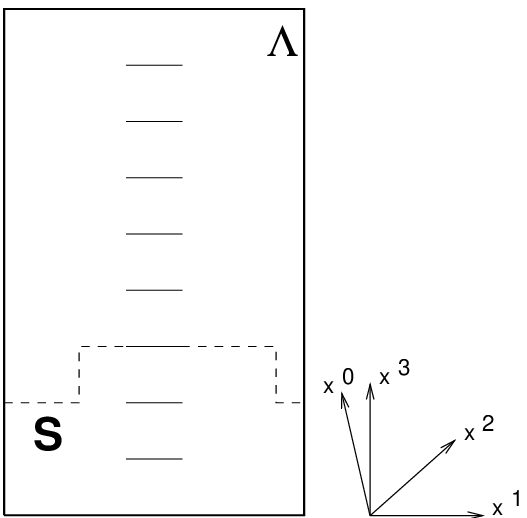}\hfill\epsfbox{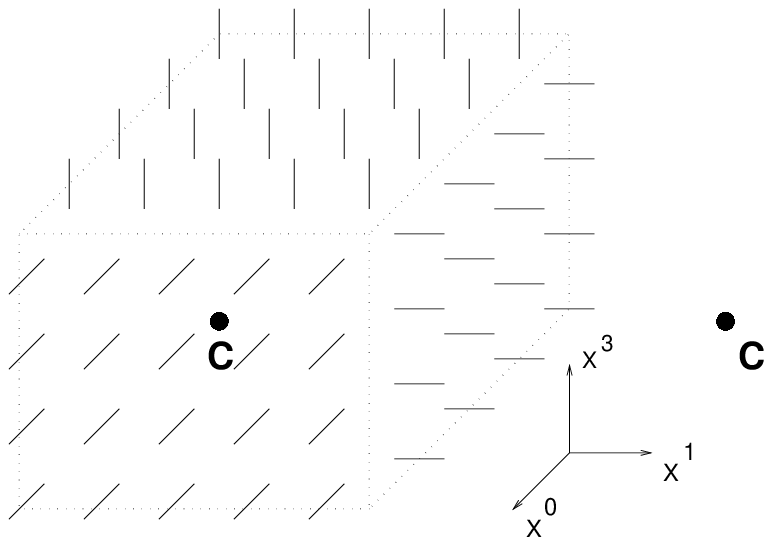}{\ }\\[0.3cm]
\hspace*{1.5cm}(a)\hfill\hbox{(b)\hspace{3.5cm}}
\caption[(a) Topologically non-trivial vortex (b) Vortex 
linked with loop C]{\label{6}(a) Topologically nontrivial 
vortex sheet (short lines) winding around periodic lattice 
(b) Vortex linked with Wilson loop $C$; 3-dimensional 
[130]-section view, with $C$ lying in the [12]-plane.}
\end{minipage}
\end{figure}

Closed Dirac plaquette sheets form the lattice 
analogs of $\pi_1(SO(3))$ vortices (Figure \ref{tckv}),  
the set of plaquettes with $\eta_p=-1$ being stacked 
over a closed 2-dimensional surface (a loop in $d=3$). 
In the illustration of Figure \ref{tckv} each 
plaquette carrying $\eta_p=-1$ protrudes in the 
$x^2$-direction out of the set of bonds shown 
distributed over a 2-dimensional 
surface placed in a 3-dimensional [013]-section.
Again, the precise location and shape of this $\eta_p=-1$ 
set of plaquettes forming the Dirac sheet
is irrelevant, it being 
`invisible' in the measure (\ref{PF1}). What is relevant is 
only the coset configuration $\{\U_b\}$ describing 
the vortex; an $SU(2)$ representative $\{U_b\}$ of 
such a coset configuration will then 
contain somewhere on $\Lambda$ a Dirac sheet, which 
may be moved around at will by a change of representatives.  
(Equivalently, the presence of the vortex can be 
characterized in terms of the $\U_b$'s only - see below). 
Note that, since $Z(2)$ (or generally $Z(N)$) flux 
is conserved only $mod\ 2$ ($mod\ N$), 
this implies that a vortex such as in Figure \ref{tckv} is 
`unstable' unless it is topologically nontrivial with 
respect to the lattice $\Lambda$ or an externally introduced 
source in $\Lambda$, such as a Wilson loop. Indeed, by a
change of representatives, the sheet may 
always be collapsed to a point 
annihilating the $Z(2)$ flux, unless there is a topological 
obstruction. Thus on a lattice with periodic boundary 
conditions, i.e. the topology of the torus, a vortex as 
in Figure \ref{tckv} may become topologically stable 
by wrapping completely around the lattice 
in the $x^3,\:x^0$-directions as shown schematically 
in Figure \ref{6}(a). Here, short light lines represent a 
Dirac sheet of $\eta_p=-1$ plaquettes, each in a 
[12]-plane, stacked along the $x^3,\:x^0$-directions around 
the periodic lattice. Such a topologically nontrivial 
closed sheet can be moved or distorted by a change 
representative, but not removed.  
Every representative of the relevant $\{\U_b\}$ 
vortex configuration has then an irremovable sheet 
of $\eta_p=-1$ plaquettes signaling the trapped 
$Z(2)$ flux of a (an odd number of) topologically nontrivial 
vortex (vortices). A characterization directly in terms of 
the $\U_b$'s  is given by the quantity $\eta_S$ defined as in 
(\ref{S}) but with $S$ now any closed topologically nontrivial 
surface winding around the lattice in the 
$x^1,\:x^2$-directions (dashed line in Figure \ref{6}(a)).
Then, clearly, $\eta_S$ is only a function 
of the cosets $\{\U_b\}$, and $\eta_S(\U)= -1(+1)$ signifies 
the presence of an odd (even) number of vortices 
winding around the lattice in the $x^3,\:x^0$-directions 
normal to $S$. An analogous description 
applies to $\pi_1(SO(3))$ vortices nontrivially linked 
with the Wilson loop (Figure \ref{6}(b)) discussed below. 

As already noted, the $Z(2)$ thin vortices are necessarily 
localized in thickness to one lattice spacing,  
and have a direct action cost proportional to their 
area (length in $d=3$) along which $\sigma_p=-1$. Long thin 
vortices are then heavily 
suppressed at large $\beta$ as the $\sigma$ $Z(2)$ variables 
are progressively frozen out. Indeed, it can be shown that 
the probability of exciting $\sigma_p=-1$ on a plaquette is 
exponentially suppressed with large $\beta$.\footnote{This 
is proven nonperturbatively by 'chessboard estimates' 
\cite{F}.}
Only short thin vortices remain then 
with (exponentially in $\beta$) small probability. 
This probability can actually depend on the choice 
of the lattice action.\footnote{Thus the density 
of the thin $Z(2)$ vortices, and/or $Z(2)$ monopoles, 
may be enhanced or further suppressed by various 
short-distance modifications of the original 
lattice (Wilson) action in (\ref{PF}). Common 
modifications in the literature involve the addition 
of chemical potentials, the MP and the ``positive 
plaquette'' models, and models introducing new $Z(2)$ degrees 
of freedom in addition to the $\sigma_p$'s\label{mods}.}
This dependence on the short 
distance structure is, of course, precisely 
a statement of the fact that thin vortices are 
thin. 

In contrast, the (lattice analog of the) 
$\pi_1(SO(3))$ vortices are not necessarily localized, 
and do not have a direct action cost proportional 
to their sheet area. Indeed, smooth $\{\U_b\}$ vortex 
configurations are easily constructed such that 
the local plaquette action cost can be 
made arbitrarily small by making the vortex sufficiently 
spread out \cite{Y,T}. Any $SU(2)$ representative 
$\{U_b\}$ of the $\{\U_b\}$ configuration for such 
a spread-out vortex will of course have a Dirac sheet 
plaquette set on which $\eta_p=-1$, as discussed above, 
while $\eta_p=1$ everywhere else; 
but in such a manner that one still has  
$|{\rm tr}U_p|\simeq 1$ everywhere. 
The exact location of the Dirac sheet is in fact 
irrelevant since it can be moved by changing representative 
which does not affect $|{\rm tr}U_p|$. 
The point is, of course, that 
the action depends only on $|{\rm tr}U_p|$, i.e. $\{\U_b\}$. 
These `thick' vortices are thus not directly suppressed 
at large  $\beta$. In fact, long thick vortices winding 
around the lattice  or a large Wilson loop can exist at 
arbitrarily weak coupling by being sufficiently thick in the 
directions transverse to their (topologically nontrivially 
linked) Dirac sheet. The same holds true for long hybrid 
vortices with a long thick vortex section, and a short 
(say, one plaquette long) thin vortex section  appearing 
as a localized `defect' incurring only a local cost in 
action \cite{T,T1}. These long hybrid vortices may 
be simply viewed as `punctured' thick vortices, the size of 
the `hole', where the $\sigma_p$ $Z(2)$ variables are 
excited, being suppressed, hence small at large $\beta$. 
 
Such very long thick vortices (whole or punctured) 
can then have a very disordering long-distance effect 
at weak coupling as the bond variables $\U_b$, over 
sufficiently large scales, vary smoothly over large 
parts of the $SU(2)/Z(2)$ group with very little local 
action cost. Their presence appears in fact to be the 
necessary condition for confinement at weak coupling as 
we discuss in the next section. 

It is also interesting to view thick vortices in the 
$d$-dimensional theory from a $(d-1) + 1$-dimensional 
perspective by singling out the `time' direction 
 \cite{T}. The $d$-dim 
gauge theory may be viewed (in Kaluza-Klein (KK) fashion) 
as a $(d-1)$-dim gauge theory coupled to a Higgs field. 
A thick vortex may then be viewed within a $(d-1)$-dim 
slice as a `monopole'-`antimonopole' pair. 
Here the `monopole` has {\it two} units ($mod\ 2$) of 
flux (and hence two $\eta$-strings emanating from it),  
since it may be considered as put together out of two 
of our $Z(2)$-monopoles and is trivial under $\pi_1(SO(3))$. 
Within a given $(d-1)$-dim slice, however, it may be 
characterized by using the `Higgs' field of the KK 
dimensional reduction to define a homotopy $\pi_2(SO(3)/U(1))$ 
group. These `monopoles' appear then as lattice analogs 
of the `t Hooft-Polyakov monopole in $(d-1)$ dimensions.  
In an appropriate gauge these correspond to the `monopoles'  
of the Abelian projection. This is, of course, not 
a fundamental, gauge invariant description of the physical 
picture which is that of the $\pi_1(SO(3))$ vortices. 
Still, it may be used as a basis for an approximate 
computational scheme for obtaining some estimate on the Wilson 
loop at weak coupling \cite{T}. Some numerical investigation of 
this picture has recently been reported in \cite{G}.

\section{The Wilson loop and Vortices}

\begin{figure}[t]
\begin{minipage}{155mm}
{\ }\hfill\epsfbox{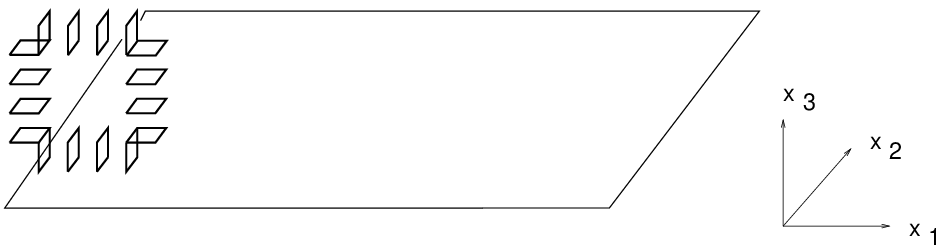}\hfill{\ }\\
\hspace*{6cm} (a) \hfill\par
{\ }\epsfbox{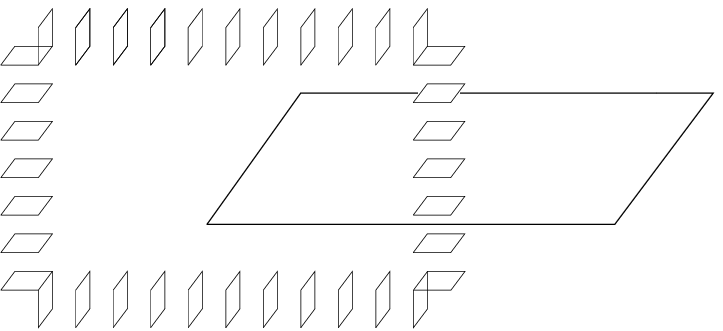}\hfill\epsfbox{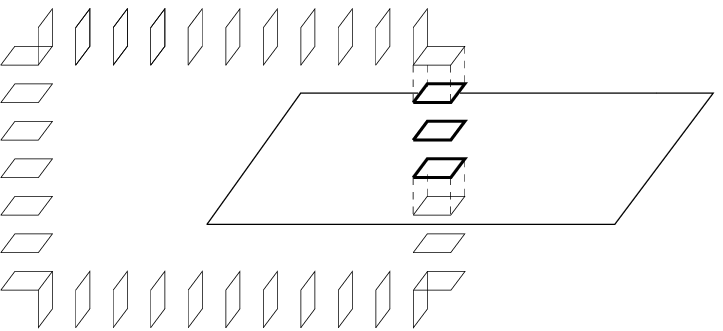}{\ }\\[0.3cm]
\hspace*{2cm} (b) \hfill \hbox{(c)\hspace{3.5cm}}\\
\caption[Vortex - Wilson loop linkage]{\label{l}Wilson  
loop linked with: (a) thin vortex, (b) thick 
vortex, (c) hybrid vortex; 3-dimensional [123]-section 
view (cp. figure \ref{6}(b)).}
\end{minipage}
\end{figure}

Having identified the various types of vortices that 
occur in the $SU(2)$ LGT, the expression (\ref{W1}) for 
the Wilson loop is seen to have a rather transparent 
physical meaning. It makes explicit the interaction of 
the loop with vortices. Let us write (\ref{W1}) 
succinctly as
\beq
W[C] = \Big\langle \;{\rm tr}U[C]\,\eta_S\,\sigma_S\;\Big 
\rangle_{SO(3)\,\cup\, Z(2)} 
\quad ,\label{W2}
\eeq
where the expectation on the rhs is taken in the measure 
(\ref{PF1}). 
As noted above, (\ref{W2}) decomposes the Wilson loop 
operator into a $Z(2)$ part $\sigma_S$ and an $SO(3)$ 
part ${\rm tr}U[C]\,\eta_S = |{\rm tr}U[C]| \;{\rm sign}\,
({\rm tr}U[C]\,\eta_S)$. It is crucial 
that the expectation (\ref{W2}) does not depend on the 
choice of the surface S spanning the loop. Then, for 
a given $\{\{\U_b\},\; \{\sigma_p\}\}$ configuration on 
the lattice: 
\begin{itemize}
\item If $\sigma_S=-1$ for {\it every} choice of the 
spanning surface $S$, a thin vortex, or an odd 
number of thin vortices, is nontrivially linked with 
the loop $C$ (Figure \ref{l}(a)). Conversely, 
$\sigma_S=1$ for every $S$ signifies an even number 
(including zero) of thin vortices linked with $C$. 
\item If $\mbox{sign}\,({\rm tr}U[C]\,\eta_S)=-1$ for 
{\it every} choice of the spanning surface $S$, a thick 
vortex, or an odd number of thick vortices, is nontrivially 
linked with the loop $C$ (Figure \ref{l}(b)). 
Conversely, ${\rm sign}\,({\rm tr}U[C]\,\eta_S)=1$ for every 
$S$ signifies an even number (including zero) of thick 
vortices linked with $C$. 
\item If neither $\sigma_S=-1$, nor ${\rm sign}\,({\rm tr}U[C]
\,\eta_S)=-1$ for {\it every} choice of $S$, but 
$\sigma_S\;{\rm sign}\,({\rm tr}U[C]\,\eta_S)=-1$  
for {\it every} choice of $S$, a hybrid vortex, or an odd 
number of hybrid vortices, is nontrivially linked 
with $C$. (Figure \ref{l}(c)). Conversely, 
$\sigma_S\;{\rm sign}\,({\rm tr}U[C]\,\eta_S)=1$ for every $S$ 
signifies that an even number (including zero) of hybrid 
vortices is linked with $C$. 
\end{itemize}

Thus the passage to the variables $\U_b,\;\sigma_p$ 
exhibits the Wilson operator explicitly as a vortex 
counter. 
The fluctuation of the operator between positive and 
negative values is entirely due to the presence of 
vortices linking with the loop. The expectation of this 
fluctuation is what essentially determines then the 
behavior of the Wilson loop. 

The physically inessential $|{\rm tr}U[C]|$, which 
contributes only a perimeter effect, can in fact be 
eliminated from expression (\ref{W1}) by switching 
from the Wilson loop to the electric-flux free energy 
order parameter $F_{el}$ \cite{tH} defined on a lattice 
with periodic boundary conditions in all directions 
($\Lambda=T^d$). Expressed in our 
variables (\ref{PF1}), the electric-flux free-energy 
is given simply by: 
\beq
\exp\:(-F_{el}) = \Big\langle\;\eta_S\,\sigma_S\;\Big
\rangle_{SO(3)\,\cup\, Z(2)}\quad ,  
\label{Fel}
\eeq   
where $S$ is any 2-dimensional closed topologically 
nontrivial surface winding around the lattice in two 
given directions as in Figure 5. (The expectation does 
not depend on the specific choice of $S$.)
By the above enumeration, (\ref{Fel}) shows that 
the electric-flux free energy operator is {\it nothing but} 
a vortex counter for the various types of vortices winding 
around the lattice (in the directions perpendicular to $S$). 

Let us return to the consideration of the Wilson loop expectation 
(\ref{W2}). At large $\beta$, long thin vortices, having 
large vortex sheet area $A$, disappear as they incur an 
action cost proportional to $\beta A$. Only a dilute gas 
of short thin vortices remains. For a large Wilson loop, 
short vortices can link with it only along the loop 
perimeter (as depicted for the thin vortex in Figure
\ref{l}(a)). At large $\beta$, therefore, thin 
vortices can only contribute at most to a length-law piece 
in the expectation (\ref{W1}). This in fact can be proven 
rigorously; it is equivalent to the statement that a pure 
$Z(2)$ theory is in a Higgs phase at large $\beta$.  

Long thick vortices, on the other hand, are not directly 
suppressed by the action at large $\beta$. They may therefore 
link with a large loop anywhere over the area enclosed by 
the loop (as in Figure \ref{l}(b)). This may then lead 
to area-law behavior for the 
expectation, provided that the class of thick vortex 
configurations contributes at large $\beta$ with a finite 
measure in the path integral sum. The same holds for long 
hybrid vortices, i.e. long thick vortices `punctured' by 
a small  monopole loop forming the boundary of a short 
thin vortex segment (Fig.\ \ref{l}(c)). Indeed, 
long $Z(2)$ monopole loops are spanned by correspondingly 
large thin vortex sheets and  
suppressed at large $\beta$; but a dilute gas of 
short monopole loops survives at any finite $\beta$. (The 
shortest possible loop is due to the excitation of 
$\sigma_p=-1$  on a single plaquette $p$, forming a 
one-plaquette-long thin segment, and giving a 
$Z(2)$ monopole loop consisting of the $2(d-2)$ cubes 
sharing this $p$ on their boundary and hence having 
$\sigma_c=-1$.) In the absence of an artificial suppression
of negative plaquettes (imposed, for example, by some 
modification of the action), this dilute gas of short Z(2) 
monopole loops can be used to tag hybrid vortices and estimate
their contribution to the Wilson loop \cite{T1}.

One may modify the theory to exclude all 
$Z(2)$ monopoles and hence all hybrid vortices by inserting 
in the measure (\ref{PF1}) the constraint
\beq 
 \prod_c\;\delta[\sigma_c] \qquad.\label{c1} 
\eeq
This is the MP model \cite{MP}.\footnote{Note that the 
solution to the constraint (\ref{c1}), which is equivalent 
to requiring $\eta_c=1$ on all cubes, is given by 
\[\sigma_p = \prod_{b\in p}\;\gamma_b \quad,\]
where $\gamma_b$ are $Z(2)$ bond variables, i.e. the $Z(2)$ 
system in (\ref{PF1}) becomes exactly a (Wilson) $Z(2)$ LGT.  
This is as one would expect: in the absence of $Z(2)$ 
monopoles, only closed thin vortices are allowed as 
excitations of the $\sigma_p$'s, which is 
the case in a pure $Z(2)$ LGT.} Confinement at large 
$\beta$ in the MP model must then come from the thick 
vortices. Alternatively, one may instead eliminate all 
thick closed vortices linking with a given Wilson loop by 
inserting in the measure in the expectation (\ref{W2}) 
the constraint 
\beq 
\theta\,\big[\:{\rm tr}U[C]\:\eta_S \:\big] \qquad.
\label{c2}  
\eeq 
for any one particular surface $S$ spanning the loop.
Similarly, thick vortices winding around a periodic 
lattice may be excluded from the theory by 
inserting the constraint 
\beq 
\delta [\,\eta_S\,]  \qquad,\label{c3}
\eeq 
for any particular closed topologically nontrivial 
surface $S$ running through the lattice in two given 
directions (cp. Figure \ref{6}(a)): (\ref{c3}) eliminates 
all vortices winding in the (d-2) directions normal to 
$S$. In the presence of the constraints (\ref{c2}) or 
(\ref{c3}), confining behavior for (\ref{W2}), or 
(\ref{Fel}), respectively, at large $\beta$ can then 
come only from the hybrid vortices. This is the 
approach taken in \cite{T1}. 

Consider now inserting {\it both} the constraints 
(\ref{c1}) and (\ref{c2}), resp.\ (\ref{c3}) in the 
measure, thus eliminating both all hybrid vortices 
{\it and}  all thick vortices winding through 
the Wilson loop, resp.\ the lattice. The form of the 
expectation (\ref{W2}), resp.\ (\ref{Fel}), now 
immediately suggests that confining behavior 
at large $\beta$ is lost, i.e. that 
{\it the presence of thick or hybrid vortices is the 
necessary condition for confinement to occur at weak 
coupling.} In the case of the electric-flux free 
energy, eq. (\ref{Fel}), a mathematically rigorous 
proof of this fact was given in \cite{Y} some time ago. 
The physical implications of this result for 
discussions of `mechanisms of confinement' appears not 
to have been widely appreciated. 
It would clearly be important to have the corresponding proof 
for the case of the Wilson loop.\footnote{Asymptotically 
large Wilson loops are physically essentially 
equivalent to the electric-flux free energy order 
parameter. In general, however, 
the string tension derived from the electric-flux free 
energy has only been rigorously shown to form a lower 
bound on the Wilson loop string tension \cite{TY}. 
Thus confining behavior for $F_{el}$ implies confining 
behavior for the Wilson loop, but not, necessarily, 
the converse. In any case, as $F_{el}$ is an order 
parameter that refers to the entire lattice, it is  
important to obtain the proof corresponding 
to the result in \cite{Y} also for a large but 
finite Wilson loop as the lattice is taken to the 
thermodynamic limit.} Unfortunately, the proof in \cite{Y} 
does not immediately extend to the Wilson loop case. 
We will address this question elsewhere.

\section{Vortex contribution to heavy-quark potential
 - Numerical results}  

As we saw in the previous section, the sign fluctuation of
the Wilson loop operator is determined by its interaction 
with vortices. (In fact, in the case of the electric-flux 
free-energy, this interaction sign constitutes the entire 
operator.) This then allows one 
to directly examine the vortex contribution to the Wilson 
loop. We simply replace the value of the 
Wilson loop operator by its sign and consider the 
expectation
\beq
E[C] \equiv \Big\langle \;\mbox{sign} \left(\,{\rm tr}U[C]\,
                                   \right)\;\Big \rangle = 
            \Big\langle \;\mbox{sign}\left(\,{\rm tr}U[C]\,
                       \eta_S \right) \sigma_S\;
                       \Big\rangle_{SO(3)\,\cup\, Z(2)}.  
          \label{E}
\eeq
The expectation (\ref{E}) is the vortex count expectation 
value as discussed above.  
In the following we wish to compare the string tension 
extracted from the full Wilson loop expectations (equations 
(\ref{W}) and (\ref{W1}) ) with the string tension obtained from
the expectation of the sign of the Wilson loops defined by equation
(\ref{E}).

In all our measurements we extracted the heavy quark potential 
from timelike Wilson loops using the method and the code 
of Ref.\ \cite{Heller}. We computed both on axis and off axis
loops and the effective potential for different time extensions
$T$ was obtained as
\beq
 V(R,T) = - \ln \frac{W(R,T+1)}{W(R,T)}.
\eeq
In principle the heavy quark potential is the $T \rightarrow \infty$
limit of $V(R,T)$. In the following we always display the
effective potential for a time extent where it has already
reached a good plateau. Typically with our values of the
coupling this already happens at T equals a few lattice spacings. 

\begin{figure}[htb!]
\begin{center}
\vskip 10mm
\leavevmode
\epsfxsize=110mm
\epsfbox{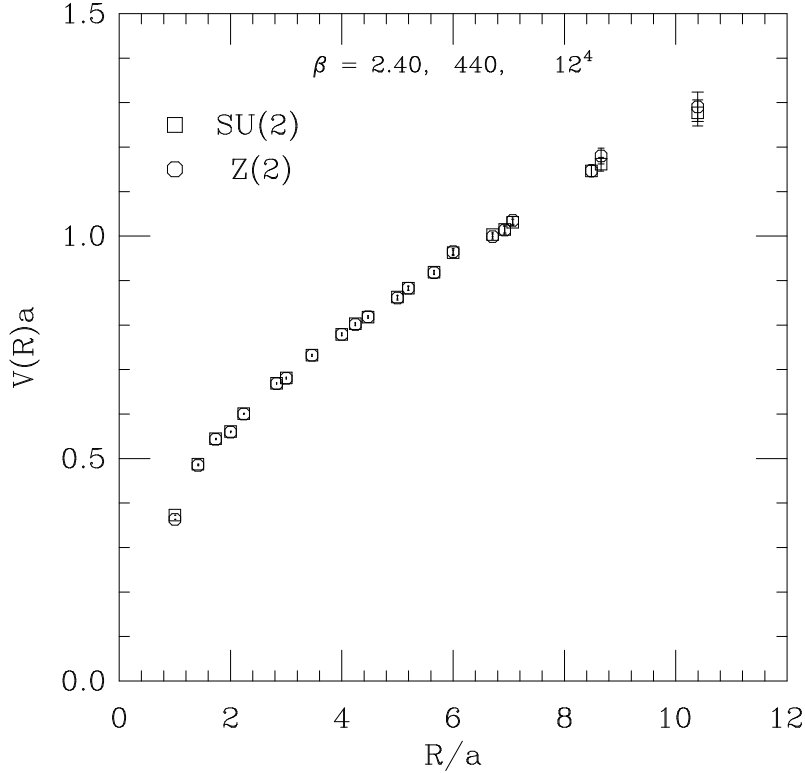}
\end{center}
\caption{The heavy quark potential measured on an ensemble of 
440 $12^4$ configurations generated at Wilson $\beta=2.4$. Squares
represent the potential obtained from Wilson loop averages, the
octagons come from the sign averages.}
\label{fig:pot12_b2.4}
\end{figure}

\begin{figure}[htb!]
\begin{center}
\vskip 10mm
\leavevmode
\epsfxsize=110mm
\epsfbox{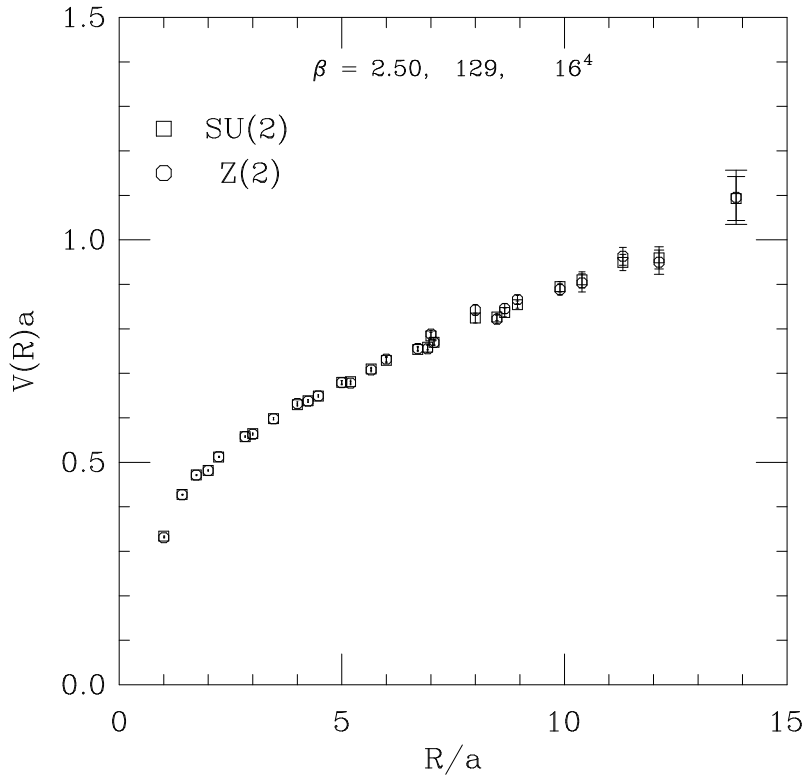}
\end{center}
\caption{The heavy quark potential measured on an ensemble of 
129 $16^4$ configurations generated at Wilson $\beta=2.5$. Squares
represent the potential obtained from Wilson loop averages, the
octagons come from the sign averages.}
\label{fig:pot16_b2.5}
\end{figure}

At first we used 
two ensembles of configurations generated with the Wilson 
action at $\beta=2.4$ and 2.5 where the lattice spacing is
$a=0.12$ fm and 0.085 fm respectively. Our results are presented
in Figs. (\ref{fig:pot12_b2.4}) and (\ref{fig:pot16_b2.5}). It is  
striking that the full Wilson loops and just their signs --- the
vortex expectations --- give exactly the same heavy quark 
potential including the short-distance behavior and even the
constant. We emphasize that we have not even shifted the
two potentials by a constant, Figs.\ (\ref{fig:pot12_b2.4}) and 
(\ref{fig:pot16_b2.5}) show the ``raw'' data without any further
manipulation.

The remarkable coincidence of the potentials computed in this
way shows that the sign of the Wilson loop, i.e.\ the number of
vortices (modulo 2) linking with it, contains all the important physics.
The short-distance agreement of the potentials can be explained
by noting that the sign expectation contains {\it all} the vortices
including thick ones $(\eta)$ and thin ones $(\sigma)$. 
The latter are important at short distances. Furthermore,
as we saw in the previous section, thin vortices affect the
perimeter-law term in the Wilson loop of any size. This in turn
contributes to the constant term in the potential. The fact
that even this constant is the same for the two potentials also
shows that they contain the same contribution from thin vortices.

At this point one could ask how robust this picture is, in
particular how sensitive it is to the physically
unimportant short distance details of the configurations. 
This can be checked either by modifying the action or by taking
the Monte Carlo generated configurations and performing some local
smoothing on them which does not change the long-distance physical 
features. If the potential extracted from the sign of the Wilson loops
is really equivalent to the full potential then their agreement at long 
distances should persist on the modified configurations. This is a 
very stringent test which has already been performed in the case of
Abelian dominance. There it turned out that while on the original 
configurations the Abelian string tension agreed with the full SU(2)
string tension to within 8\%, after smoothing the difference increased to
about 30\% \cite{Zsolt}. Similar results have been obtained with 
cooling in Ref.\ \cite{Hart}.
\begin{figure}[htb!]
\begin{center}
\vskip 10mm
\leavevmode
\epsfxsize=110mm
\epsfbox{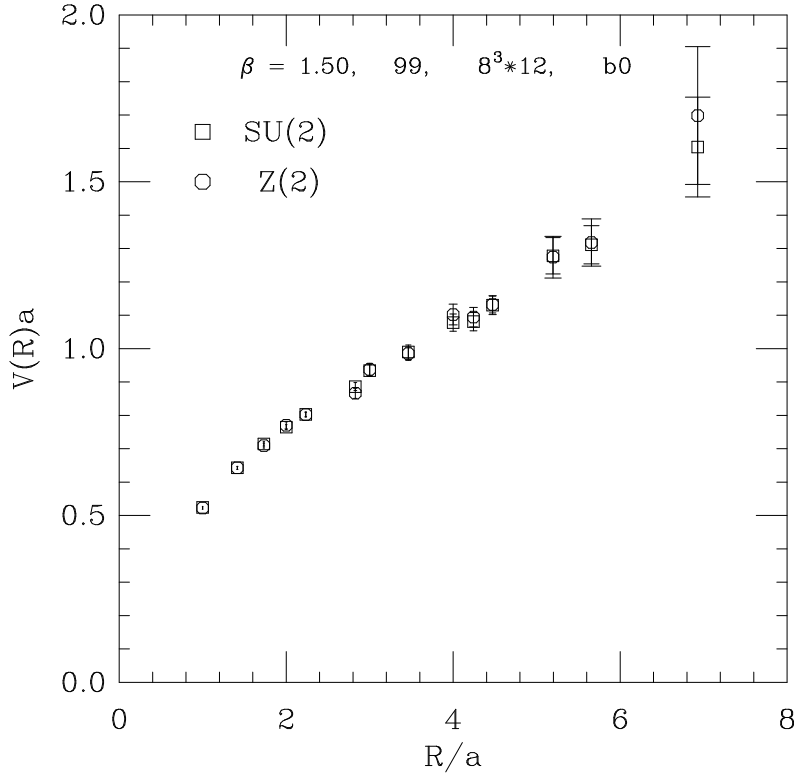}
\end{center}
\caption{The heavy quark potential measured on an ensemble of 
100 $8^3\times12$ configurations generated with a fixed point action
(lattice spacing $a=0.14$ fm.). Squares
represent the potential obtained from Wilson loop averages, the
octagons come from the sign averages.}
\label{fig:pot812_b1.5_b0}
\end{figure}
In the present case at first we repeated 
the measurement of the full and ``sign'' potentials
using the fixed point action of Ref.\ \cite{DG} at lattice spacing
$a=0.14$ fm. The results presented in Fig.\ (\ref{fig:pot812_b1.5_b0})
are very similar to the Wilson data; there is no measurable
difference between the potentials. We then performed one step of local 
smoothing on the same ensemble of configurations. This was done by the
renormalization group based smoothing introduced in Ref.\ \cite{DG}.
This local smoothing was designed to smooth only on the shortest 
distance scale, leaving all the long-distance physical features --- most
notably the string tension --- unchanged.
On the smoothed configurations the two potentials were measured again.
Comparing the potentials obtained on the smoothed configurations (Fig.\
(\ref{fig:pot812_b1.5_b1})) one can see that for distances $R \geq 2$
(in lattice units) they agree but for $R \leq 2$ the potential obtained
from the signs is systematically below the full potential. This means that
the smoothing destroyed a significant number of thin vortices and on this
short distance scale thin vortices no longer dominate the the potential. 
On the other hand thicker vortices could not be destroyed by a local 
smoothing and the long-distance features are thus preserved. 
\begin{figure}[htb!]
\begin{center}
\vskip 10mm
\leavevmode
\epsfxsize=110mm
\epsfbox{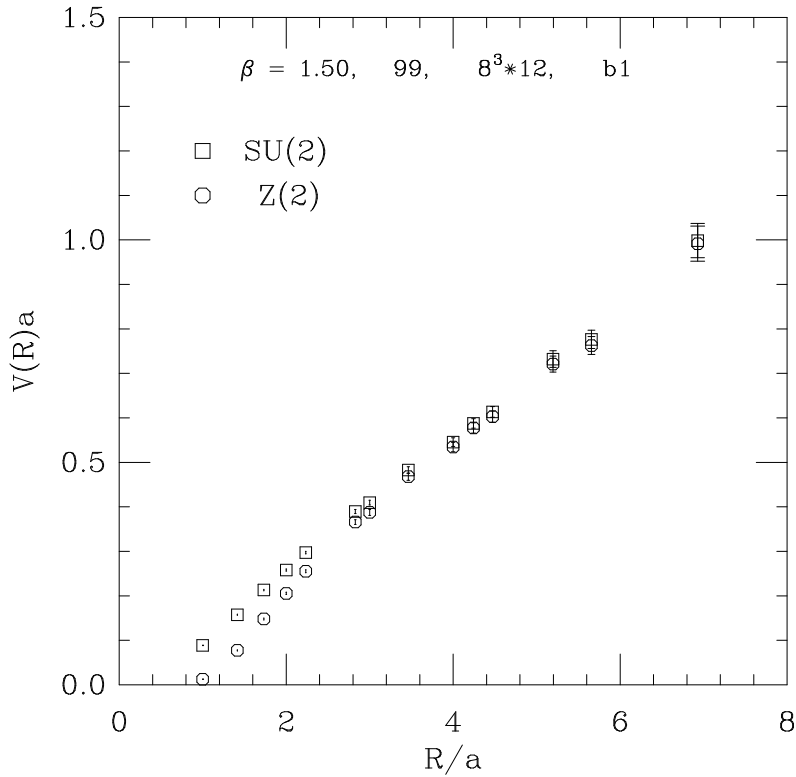}
\end{center}
\caption{The heavy quark potential on the same ensemble as Fig.\ 
\ref{fig:pot812_b1.5_b0} but measured after one smoothing step. Squares
represent the potential obtained from Wilson loop averages, the
octagons come from the sign averages.}
\label{fig:pot812_b1.5_b1}
\end{figure} 
In this context we note that exactly the same type of behavior would
be expected from the positive plaquette model, in which the plaquettes
are constrained to be non-negative. This constraint does not allow the
formation of thin vortices but vortices thicker than one plaquette are 
not affected significantly.

\begin{figure}[htb!]
\begin{center}
\vskip 10mm
\leavevmode
\epsfxsize=110mm
\epsfbox{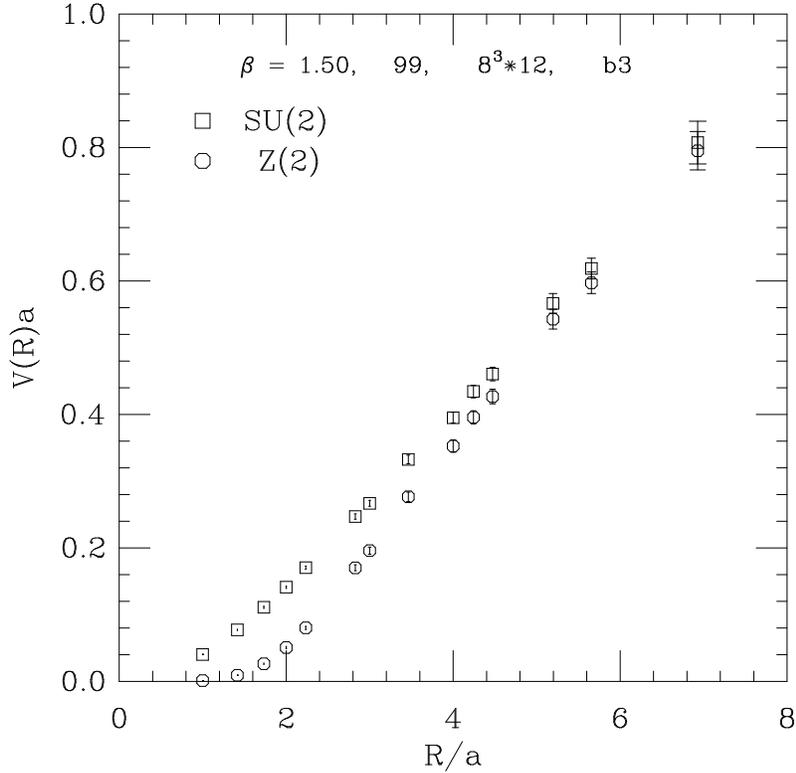}
\end{center}
\caption{The same as Fig.\
\ref{fig:pot812_b1.5_b0} but measured after 3 smoothing steps. Squares
represent the potential obtained from Wilson loop averages, the
octagons are obtained from Wilson loop sign averages.}
\label{fig:pot812_b1.5_b3}
\end{figure}

Finally we repeated the comparison of the full and the sign potential
after an additional two smoothing steps were performed (Fig.\ 
\ref{fig:pot812_b1.5_b3}). As a result of further smoothing 
the short distance disagreement of the potentials extended to 
a bit longer distances but the asymptotic string tension is not
affected. This is consistent with our expectations that as more and
more smoothing is performed, vortices of larger size are also destroyed.
For a fixed number of smoothing steps, however, there is always a scale
beyond which thick vortices remain intact. Beyond this scale one
effectively has the same physical situation as before the smoothing.
This may be viewed as being on a fictitious coarser lattice with the
lattice spacing set by this scale and with thin and thick vortices 
relative to this scale. Thus the vortex contribution to the asymptotic 
string tension is not affected by smoothing.

\section{Conclusions}

We presented a picture of the QCD vacuum by identifying
the gauge field excitations that can disorder the system on 
large distance scales and can thus lead to confinement even 
at weak coupling. The relevant excitations are thick spread
out center vortices that make the sign of large Wilson loops 
fluctuate considerably. The vortices are extended  
objects that cost very little in local action but have a 
long-range disordering effect. As opposed to thin
vortices which gradually freeze out when the coupling is lowered, 
the thick vortices are expected to survive at arbitrarily 
weak couplings.

We tested numerically how the vortices affect the Wilson loop expectations
and the deduced heavy quark potential. In the SU(2) case vortices linking 
with the Wilson loop are responsible for the fluctuation of its sign. Therefore
we compared the heavy-quark potential extracted from full Wilson loops
with the potential extracted from the expectation of the sign of Wilson loops.
The measurements were performed with the Wilson action at two different 
couplings as well as with a perfect action. In all three cases the two 
potentials completely agreed even for small distances.

To check the universality of this picture we repeated the same test on
an ensemble of locally smoothed configurations. The agreement of the long-distance
part of the potentials persisted after the smoothing. This shows
that all the relevant long-distance physical properties are encoded 
in the fluctuation of the sign of the Wilson loops which in turn is governed
by the vortices linking with it.

\end{document}